\documentclass[aps,prc,preprint,amsmath,amssymb,superscriptaddress,showpacs]{revtex4-1}
\usepackage{graphicx}  
\usepackage{float}
\newcommand{\amev}{\textit{A} MeV}
\begin{document}


\title{Isospin influence on dynamical production of Intermediate Mass Fragments  at Fermi Energies
}

\author{P.~Russotto}\email{russotto@lns.infn.it}\affiliation{INFN, Laboratori Nazionali del Sud, Catania, Italy}

\author{E.~De~Filippo}\affiliation{INFN, Sezione di Catania, Catania, Italy}

\author{E.V.~Pagano}\affiliation{INFN, Laboratori Nazionali del Sud, Catania, Italy}
\affiliation{Dipartimento di Fisica e Astronomia, Univ. di Catania, Catania, Italy}

\author{L.~Acosta}\affiliation{INFN, Laboratori Nazionali del Sud, Catania, Italy}
\affiliation{Instituto de F\'isica, Universidad Nacional Aut\'onoma de M\'exico, Mexico.}

\author{L.~Auditore}\affiliation{INFN, Sezione di Catania, Catania, Italy} 
\affiliation{Dipartimento di Scienze MIFT, Univ. di Messina, Messina, Italy}


\author{T.~Cap}\affiliation{National Centre for Nuclear Research, Otwock-\'Swierk, Poland}

\author{G.~Cardella}\affiliation{INFN, Sezione di Catania, Catania, Italy}



\author{S.~De Luca}\affiliation{INFN, Sezione di Catania, Catania, Italy} 
\affiliation{Dipartimento di Scienze MIFT, Univ. di Messina, Messina, Italy}



\author{B.~Gnoffo}\affiliation{INFN, Sezione di Catania, Catania, Italy}
\affiliation{Dipartimento di Fisica e Astronomia, Univ. di Catania, Catania, Italy}

\author{G.~Lanzalone}\affiliation{INFN, Laboratori Nazionali del Sud, Catania, Italy}
\affiliation{Universit\`a di Enna ``Kore'', Enna, Italy}

\author{I.~Lombardo}\affiliation{INFN, Sezione di Catania, Catania, Italy}
\affiliation{Dip. di Fisica "E. Pancini", Univ. di Napoli Federico II, Napoli, Italy}

\author{C.~Maiolino}\affiliation{INFN, Laboratori Nazionali del Sud, Catania, Italy}

\author{N.S.~Martorana}\affiliation{INFN, Laboratori Nazionali del Sud, Catania, Italy}
\affiliation{Dipartimento di Fisica e Astronomia, Univ. di Catania, Catania, Italy}

\author{T.~Minniti}\altaffiliation{current address: STFC, Rutherford Appleton Laboratory, ISIS Facility, Harwell, OX11 0QX,
United Kingdom}
\affiliation{INFN, Sezione di Catania, Catania, Italy}

\author{S.~Norella}\affiliation{INFN, Sezione di Catania, Catania, Italy} 
\affiliation{Dipartimento di Scienze MIFT, Univ. di Messina, Messina, Italy}

\author{A.~Pagano}\affiliation{INFN, Sezione di Catania, Catania, Italy}

\author{M.~Papa}\affiliation{INFN, Sezione di Catania, Catania, Italy}

\author{E.~Piasecki}\affiliation{Heavy Ion Laboratory, University of Warsaw, Warsaw, Poland}

\author{S.~Pirrone}\affiliation{INFN, Sezione di Catania, Catania, Italy}

\author{G.~Politi}\affiliation{INFN, Sezione di Catania, Catania, Italy}
\affiliation{Dipartimento di Fisica e Astronomia, Univ. di Catania, Catania, Italy}

\author{F.~Porto}\affiliation{INFN, Laboratori Nazionali del Sud, Catania, Italy}
\affiliation{Dipartimento di Fisica e Astronomia, Univ. di Catania, Catania, Italy}

\author{L.~Quattrocchi}\affiliation{INFN, Sezione di Catania, Catania, Italy}
\affiliation{Dipartimento di Fisica e Astronomia, Univ. di Catania, Catania, Italy}

\author{F.~Rizzo}\affiliation{INFN, Laboratori Nazionali del Sud, Catania, Italy}
\affiliation{Dipartimento di Fisica e Astronomia, Univ. di Catania, Catania, Italy}

\author{E.~Rosato}\altaffiliation{deceased}
\affiliation{INFN, Sezione di Napoli and Dipartimento di Fisica, Univ. di Napoli, Italy}

\author{K.~Siwek-Wilczy\'nska}\affiliation{Faculty of Physics, University of Warsaw, Warsaw, Poland}

\author{A.~Trifir\`o}\affiliation{INFN, Sezione di Catania, Catania, Italy} 
\affiliation{Dipartimento di Scienze MIFT, Univ. di Messina, Messina, Italy}

\author{M.~Trimarchi}\affiliation{INFN, Sezione di Catania, Catania, Italy} 
\affiliation{Dipartimento di Scienze MIFT, Univ. di Messina, Messina, Italy}

\author{G.~Verde}\altaffiliation{current address: Institut de Physique Nucl\'eaire, CNRS-IN2P3, Univ. Paris-Sud, Orsay, Universit\'e Paris-Saclay, Orsay, Cedex, France}
\affiliation{INFN, Sezione di Catania, Catania, Italy}


\author{J.~Wilczy\'nski}\altaffiliation{deceased}
\affiliation{National Centre for Nuclear Research, Otwock-\'Swierk, Poland}



\date{\today}

\begin{abstract}

The Intermediate Mass Fragments emission probability from Projectile-Like Fragment break-up in semi-peripheral reactions has been measured in collisions of $^{124}$Xe projectiles with two different targets of $^{64}$Ni and $^{64}$Zn at the laboratory energy of 35 \amev. The two colliding systems differ only for the target atomic number Z and, consequently, for the Isospin $N/Z$ ratio. An enhancement of Intermediate Mass Fragments production for the neutron rich $^{64}$Ni target, with respect to the $^{64}$Zn, is found.
In the case of one Intermediate Mass Fragment emission, the contributions of the dynamical and statistical emissions have been evaluated, showing that the increase of the effect above is due to an enhancement of the dynamical emission probability, especially for heavy IMFs (Z$\gtrsim$ 7). This proves an influence of the target Isospin on inducing the dynamical fragment production from Projectile-Like Fragment break-up. In addition, a comparison of the Xe+Ni,Zn results with the previously studied $^{112,124}Sn+^{58,64}Ni$ systems is discussed in order to investigate the influence of the projectile Isospin alone and to disentangle between Isospin effects against system-size effects on the emission probability. These comparisons suggest  that the prompt-dynamical emission is mainly ruled by the $N/Z$ content of, both, projectile and target;  for the cases here investigated, the influence of the system size on the dynamical emission probability can be excluded.
%
\end{abstract}



\maketitle

\section{INTRODUCTION}
In Heavy-Ion collisions at Fermi energies the production of Light Particles and Intermediate Mass Fragments (IMFs), here defined as fragments with atomic number $Z\geq3$, is due to different reaction mechanisms related to different time scales \cite{Pag04,Def05a,Def05b,Dit06,Pap07,Pia07,Rus10,McI10,Def12,Def14,Rod17,Jed17}.
This was seen in semi-peripheral collisions where coexistence of light IMFs emission from prompt neck-fragmentation process \cite{Pag04,Def05a,Dit06,Pap07,Pia07,Def12,Def14}, projectile-like fragments (PLF*) collinear massive break-up \cite{Boc00,Def05b,Pap07,Rus10,McI10,Rus15} and equilibrated PLF* binary fission-like emission \cite{Boc00,Rus10,Def14,Rus15} has been reported within the same range of large relative impact parameters. Thus, within a narrow range of initial kinematical nucleus-nucleus conditions of relative energy and angular momenta, the competition among different mechanisms having a large difference in the time scale,  from prompt IMFs emissions (dozens of fm/c) to equilibrated decay ($\gg 300 fm/c$), has been observed.
This stimulates for further investigations in order to elucidate the  roles played by the initial properties of the systems, the Isospin ($N/Z$) degree of freedom or the system size, in triggering the above mentioned reaction mechanisms, thus affecting their probability of occurence. These studies, investigating how the reaction mechanism path depends on basic nuclear properties, such as the Isospin content, are also important in order to shed lights on the effective in-medium interaction \cite{Sch01,Bar05,BaLi08,Oer17}.\\
Aim of this work is to show a physics case in which the Isospin degree of freedom plays a characteristic role in influencing the reaction path, affecting the occurrence of dynamical processes.\\
In fact, in a previous work \cite{Rus15}, we studied the massive IMFs emission in semi-peripheral collisions of neutron rich $^{124}$Sn+$^{64}$Ni and neutron poor $^{112}$Sn+$^{58}$Ni systems at the beam energy of 35 \amev~(REVERSE experiment). The Dynamical (prompt) IMFs emission was proved to be more probable for the neutron rich system. 
In that work emphasis was given to events with emission of only one IMF, representing the largest fraction ($\sim 60\%$) of the IMFs multiplicity distribution. As main results, we found that the IMFs statistical-equilibrated binary fission-like emission was equally probable for the two systems, while the dynamical splitting was favored in the case of the neutron rich system, especially for IMFs heavier than Z=7. Since the atomic numbers are equal (for both projectile and target) in the two Sn+Ni systems, the enhancement of the dynamical emission in the neutron rich system was mainly ascribed to the different $N/Z$ ratio, indicating an entrance channel Isospin effect, even if, at that stage, a possible influence of the mass difference (A=188 vs A=170) could not be excluded.

In order to disentangle genuine entrance channel Isospin effects from the possible dependence on the size of the two systems, we carried out a new experiment, as it was mentioned in Ref. \cite{Rus15}, named "Inverse Kinematic Isobaric Systems", using the $^{124}Xe$+$^{64}Zn$ system, that is, a projectile/target combination having the same mass as the neutron rich $^{124}Sn+^{64}Ni$ system and a $N/Z$ near to the value of the neutron poor $^{112}Sn$+$^{58}Ni$ one, at the same bombarding energy of 35 \amev; in addition also $^{124}Xe$+$^{64}Ni$ was studied. For comparison, the $N/Z$ of the new investigated systems are compared in table \ref{tab1}, with the ones from Ref. \cite{Rus15}.
\begin{table}[h]
\caption{\label{tab1}Isospin ($N/Z$) of the  systems investigated in the discussed experiments}
\begin{ruledtabular}
\begin{tabular}{*{4}{c}}
\centering   
$System$&$N/Z~Projectile$&$N/Z~Target$&$N/Z~Composite$\\ 
$^{124}Sn$+$^{64}Ni$&$1.48$&$1.29$&$1.41$\\
$^{112}Sn$+$^{58}Ni$&$1.24$&$1.07$&$1.18$\\
$^{124}Xe$+$^{64}Zn$&$1.30$&$1.13$&$1.24$\\
$^{124}Xe$+$^{64}Ni$&$1.30$&$1.29$&$1.29$\\
\end{tabular}
\end{ruledtabular}
\end{table}

\section{THE EXPERIMENT}
The data discussed in this paper have been collected at the INFN-LNS in Catania by using the $4\pi$ CHIMERA multi-detector \cite{Pag12, Def14}. A $10^{7}$ pps beam of $^{124}Xe$ bombarded $^{64}Zn$  (308 $\mu$g/cm$^{2}$) and $^{64}Ni$ (370 $\mu$g/cm$^{2}$) targets; events were registered requiring at least 2 (or 3 in a part of the experiment) charged particle hits in the CHIMERA silicon detectors. As in previous experiments \cite{Def14}, particles punching trough the silicon detectors were identified in charge, with
also isotopic identification for light fragments (Z$\leq$9),  by using the $\Delta$E-E technique. Light Charged Particles (LCPs) of
atomic numbers Z$\leq$2 were isotopically identified by the two-gates integration Pulse Shape analysis (Fast-Slow technique) of CsI(Tl)
signals \cite{Aie96}. As typical in CHIMERA detector, direct velocity measurement for ions of Z$>$2, and information on the mass number of heavy particles stopped in silicon detectors, were obtained  via the
time-of-flight (TOF) technique. 
In addition, slow moving particles stopped in the silicon detectors, with threshold energy E$\gtrsim$4.5 AMeV and emitted at $\theta_{lab}>7^o$,  were also identified in charge via Pulse Shape analysis (Energy-Rise Time correlation technique) \cite{PStec}.

 In the present analysis, differing from the work of Ref. \cite{Rus15}, the first ring of CHIMERA ($1.0^o<\theta_{lab}<1.8^o$) was not included to avoid spurious pile-up contaminants at small detection angles. Also we cannot extract cross section from this data, as instead done in Ref. \cite{Rus15}, since the measurement of the elastic scattering in the monitor detector was not allowed for technical reasons.
A small portion of the solid angle ($\sim 160~msr$, $\theta_{lab} \approx 15^o-45^o$ and $\Delta\phi_{lab}\approx 90^o$ ) was used to allocate a first prototype of the FARCOS \cite{EVP16,ACO16} detector to collect IMFs, with high angular resolution,  in order to pursue investigations on the feasibility of IMF-IMF relative momentum correlation measurements, to be exploited in future dedicated experiments \cite{CHIFAR}. This resulted in covering of some CHIMERA telescopes. FARCOS data analysis is not included in this paper and will be published separately \cite{EVP16,EVPXX}.\\


\section{DATA ANALYSIS AND RESULTS}
\begin{figure*}[h]
\includegraphics[width=1.0\textwidth]{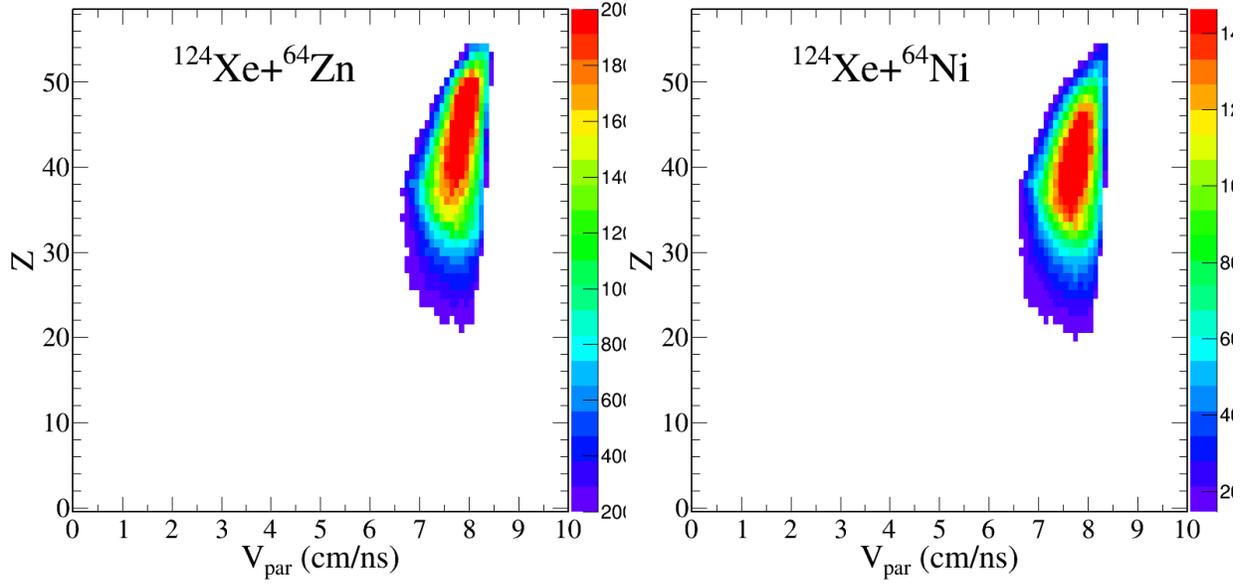}
\caption{(Color online)  Atomic number $Z$ versus parallel-to-the-beam  velocity $V_{par}$ plots for PLF, selected according to the method described in Refs. \cite{Def05a,Rus15}, in the $^{124}$Xe+$^{64}$Zn (left panel) and in the $^{124}$Xe+$^{64}$Ni systems (right panel).
\label{fig01}}
\end{figure*}

\begin{figure*}[h]
\includegraphics[width=1.0\textwidth]{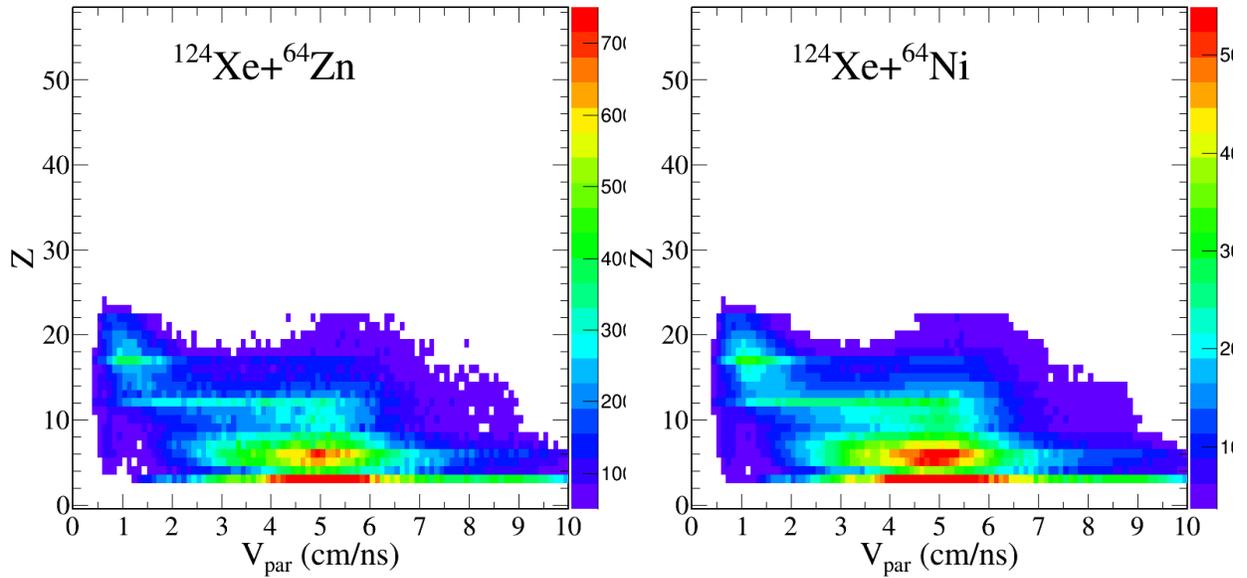}
\caption{(Color online)  Atomic number $Z$ versus parallel-to-the-beam  velocity $V_{par}$ plots for fragments detected in coincidence with the PLF shown in Fig. \ref{fig01}, in the $^{124}$Xe+$^{64}$Zn (left panel) and in the $^{124}$Xe+$^{64}$Ni systems (right panel).
\label{fig02}}
\end{figure*}

\begin{figure}[h]
\includegraphics[width=1.0\textwidth]{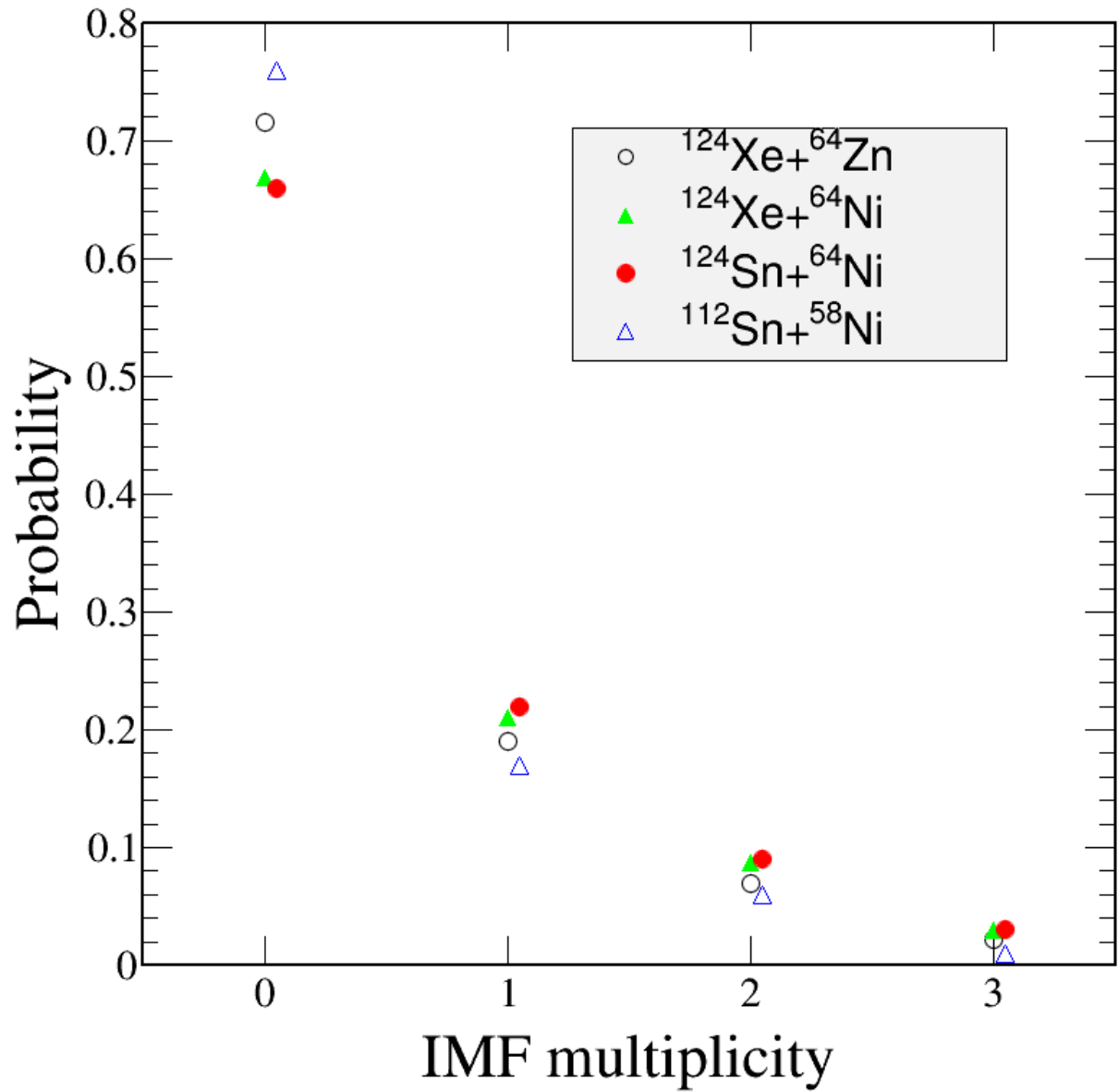}
\caption{(Color online)  Probability of different IMF multiplicities (in coincidence with PLF residues) for the $^{124}$Xe+$^{64}$Ni (empty circles) and $^{124}$Xe+$^{64}$Zn (full triangles) systems; for comparison, also the data of the $^{124}$Sn+$^{64}$Ni (full circles) and $^{112}$Sn+$^{58}$Ni (empty triangles) systems are shown. The error bars are smaller than symbols size.
\label{fig03}}
\end{figure}

In order to disentangle dynamical and statistical emissions and evaluate their probabilities, we have applied the analysis method already used in \cite{Rus15} and there fully described. Briefly, for both systems, we select semi-peripheral collisions by applying a contour gate in the $Z-V_{par}$ two-dimensional plot of detected fragments, where Z and $V_{par}$ represent, respectively, atomic number and parallel velocity component to the beam axis. The $Z-V_{par}$  correlation plots of selected PLFs, i.e., the heavy remnant of primary PLF* break-up or evaporation \cite{Not}, are shown in the two panels of Fig. \ref{fig01}, for the Zn (left panel) and Ni (right panel) targets. The $Z-V_{par}$ correlation plots of IMFs, together with target-like fragment (TLF), detected in coincidence with the selected PLF, are shown in the two panels of Fig. \ref{fig02} for the Zn and Ni targets. The similarity between the two plots proves that the calibration and identification procedures have not introduced differences in the data collected for the two different targets.\\The TLF* products were strongly suppressed by selecting only IMFs having a parallel velocity greater than 3 cm/ns. The IMF multiplicity distribution originated from the PLF*, under such a selection, is shown in Fig. \ref{fig03}; there the IMF multiplicity distribution previously measured for the two $^{124,112}Sn+^{64,58}Ni$ systems is also reported. The new data have been corrected for the efficiency of the experimental apparatus (for $\theta_{lab}>1.8^{\circ}$), including the small losses due to the FARCOS prototype coverage. However, we have not applied any correction for the missing coverage at
smaller $\theta_{lab}$ region, since we have verified that the resulting estimation of the lost efficiency will be drastically dependent on the input scheme/model used to estimate it. Evidently, it means that a part of very peripheral and low-dissipative events, with the PLF* only slightly deflected from the beam direction, have not been included in this analysis. Consequently, the previous REVERSE Sn+Ni data \cite{Rus15} shown in this paper have been filtered, and corrected for the detection efficiency, in the same way as the Xe+Ni,Zn ones.
  
As a first observation, by comparing the results obtained with Xe projectiles, we find an enhancement of IMFs emission probability in the case of the Ni neutron rich target. This means that, for isobaric targets, the IMFs emission probability from PLF* fragmentation is enhanced by a small increase of the target $N/Z$ content, i.e., 2 neutrons are replaced by 2 protons. 
In \cite{Rus15} we already investigated that effect for the two Sn+Ni systems, finding an higher IMFs emission probability for the neutron rich $^{124}Sn+^{64}Ni$ system, also reported in Fig. \ref{fig03}. There we clearly showed that this enhancement was essentially due to an increase of the dynamical IMFs emission alone, beyond the simple expectation of an higher IMFs emission in the neutron rich systems. As shown below, also in the case of Xe+Zn,Ni systems the enhancement of IMFs emission for the neutron rich target can be ascribed to an increase of the dynamical emission. In the class of semi-peripheral collisions investigated in this work, the energy/mass and angular-momentum transfers, taking place during interaction and re-separation phases, play a decisive role in driving the PLF* fate \cite{Bar04,Bar05,Pap07}.
In fact, the dynamical emission, from PLF* in this case, is essentially due or to bulk instabilities at low densities (neck-emission) taking place in the early phase of the re-separation, or to shape instabilities (dynamical fission or aligned break-up) leading the PLF* to a fast non-equilibrated fragmentation \cite{Col98,Bar04,Pap07} . It follows that these instabilities are more easily triggered in the case of interaction of projectile with a neutron rich target. This observation highlights the importance of the target $N/Z$ content alone in such early dynamical phases.\\
Furthermore, notice that by comparing the results with $^{64}Ni$ target and $^{124}Sn$-$^{124}Xe$ projectiles, an enhancement of IMFs emission probability from PLF* fragmentation in the case of the $^{124}Sn$ neutron rich projectile is also found. This effect is in agreement with the expectation of an enhancement of IMFs production in the case of a neutron rich system; however, as we will show below, this can be also connected to an increase of the dynamical emission probability in the case of a neutron rich projectile, thus, envisaging also the role of the projectile $N/Z$ content on the development of instabilities in the above mentioned early-dynamical phases.\\In order to corroborate the results here described and to shed lights on how they depends on properties of the many body microscopic interaction, simulations with dynamical transport models able to careful describe the complex processes of the projectile-target interaction and re-separation phases and to follow the full time scale path (from dynamical to equilibrated emission) of the excited PLF*, practically, up to about 1000 fm/c, are needed. Different transport models \cite{Bar04,Bar05,Pap07} are available for the simulation of the early dynamic phase but, unfortunately, they tend to be unstable beyond that phase or fail when trying to reproduce the full dynamical evolution of the PLF*, especially when the whole IMF mass spectrum (up to heavier IMFs) is needed, as it is the case discussed in this work. Overcoming of these problems of simulations will allow to clearly improve the effectiveness of studies like the one here reported.\\ 
By looking at the probability of IMF multiplicity=0 (reactions leading to PLF-TLF pair accompanied in coincidence by only few LCPs in the final state) a bigger differences is found when changing the target (from Ni to Zn) than when changing the projectile (from Xe to Sn): this suggests that, in the cases here investigate, that is a heavy projectile on a lighter target, the IMF emission probability shows a greater sensitivity on the target $N/Z$ content rather than on the projectile one.\\
Then we have selected the IMF multiplicity=1 and we have evaluated the  dynamical and statistical IMFs emission from PLF* source by applying the same criterion of Ref. \cite{Rus15}.
To better select PLF* emission (after excluding TLF* products slower than 3 cm/ns), the PLF and IMF relative velocity was limited by the constraint $V_{rel}(PLF,IMF)/V_{Viola}\leq 1.5$ , where $V_{Viola}$  represents the mutual PLF-IMF  Coulomb repulsion in a sequential $PLF^{*}\rightarrow PLF+IMF$ decay as evaluated by using the Viola systematics \cite{Vio85}.

\begin{figure}[h]
\includegraphics[width=1.0\textwidth]{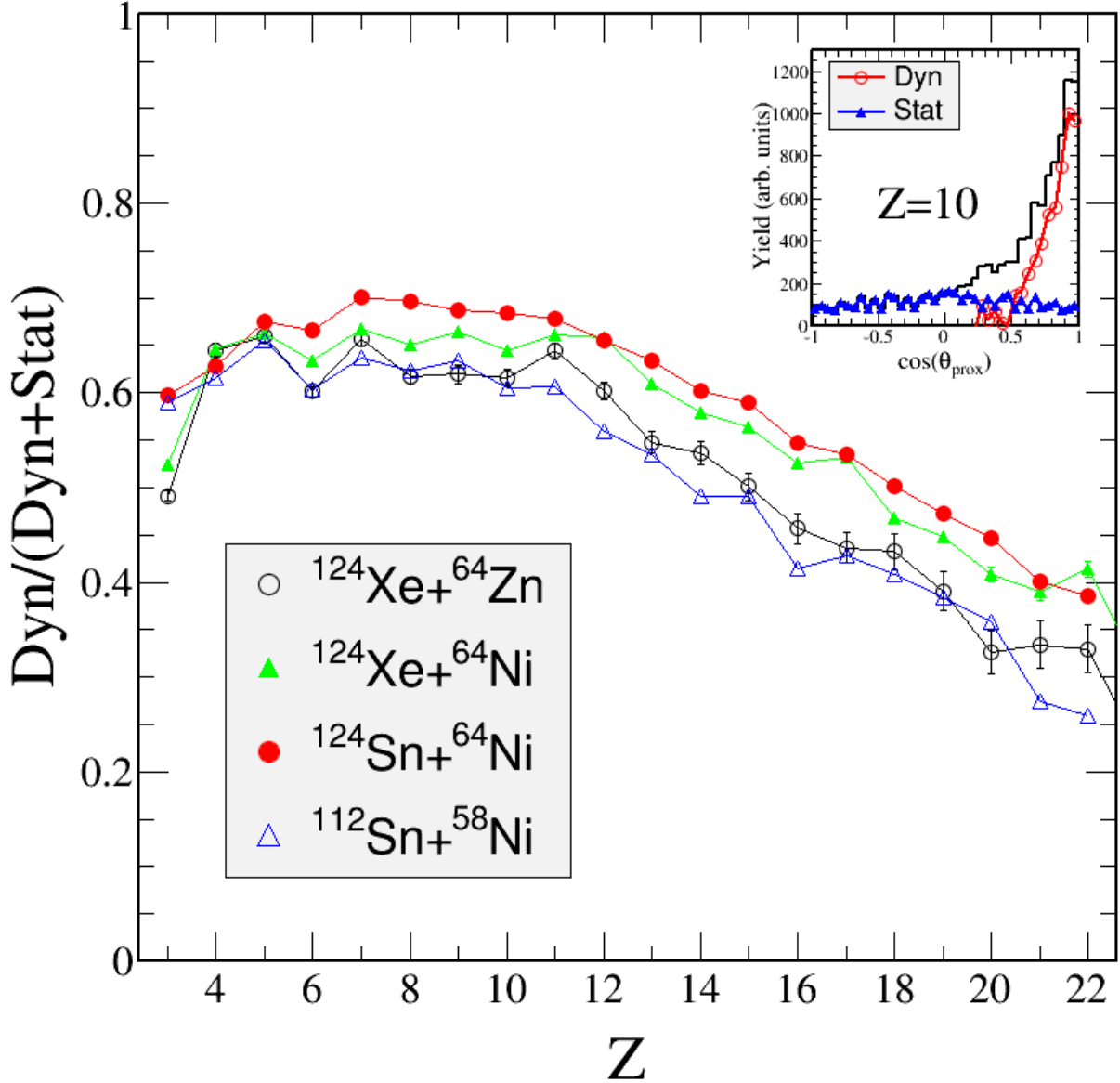}
\caption{(Color online) Ratio of the dynamical component to the total (dynamical+statistical) value plotted as
a function of the IMF atomic number $Z$, for the $^{124}$Xe+$^{64}$Zn (empty circles) and $^{124}$Xe+$^{64}$Ni (full
triangles) systems; for comparison, also the data of the $^{124}$Sn+$^{64}$Ni (full circles) and $^{112}$Sn+$^{58}$Ni (empty triangles) systems are shown. The inset shows an example of the $\cos(\theta_{prox})$ for IMFs of Z=10 (black line). The statistical contribution (full triangles) is obtained symmetrizing around $\cos(\theta_{prox})=0$ the "forward emission", i.e., the $\cos(\theta_{prox})<0$ part; the Dynamical contribution (empty circles) is then obtained by subtracting the statistical contribution from the total one, as described in \cite{Boc00}. 
\label{fig04}}
\end{figure}

For PLF-IMF pair satisfying this latter condition, the $\cos(\theta_{prox})$ distributions were evaluated, as a function of the Z of the IMFs; $\theta_{prox}$ is the angle between
the breakup  axis, oriented  from the light (IMF) to the heavy (PLF) fragment of the primary PLF* break-up, and  the  recoil  velocity  in  the  center of mass reference of the PLF*, reconstructed by the masses and velocity vectors the two PLF-IMF fragments (see \cite{Boc00} for definition). The weight of the dynamical and statistical components in the $\cos(\theta_{prox})$ distribution, disentangled using the method described in \cite{Rus15}, was simply evaluated (see, also, inset and caption of Fig. \ref{fig04}).\\
We think useful to point out at the similarity and difference of this method with the one used in Ref. \cite{Rod17,Jed17} (and other works of that collaboration), where dynamical IMFs emission has also been careful investigated. Indeed, both methods are based on the analysis of the PLF* break-up angular distributions; while in the works of Ref. \cite{Jed17} the attention is focused on the relation between the emission angle and the N/Z of the break-up products, investigating the time scale of the neutron-proton equilibration, as we partially did in ref. \cite{Def12}, in the present work and in \cite{Rus15} the angular distributions, for each Z, are used to quantitatively estimate the probabilities of dynamical and statistical emission. Hence, this is the main observable we are discussing.

The ratio of the dynamical component to the total (dynamical+statistical) contribution is shown in Fig. \ref{fig04}, as a function of the atomic number $Z$ of the IMFs. In the figure, also the ratio for the two $^{124,112}Sn+^{64,58}Ni$ systems is reported. By looking at the systems with Xe projectile, we can see an enhancement of dynamical emission for heavier IMFs having $Z\gtrsim7$ for the neutron rich Ni target. This means that, for isobaric targets, the IMFs dynamical emission probability from PLF* binary break-up is enhanced by the increase of the target $N/Z$ content. Also, by comparing the results of heavier IMFs when using the $^{64}Ni$ target, an enhancement of dynamical emission probability in the case of the $^{124}Sn$ neutron rich projectile is found. 
Also in this case, as in the discussion of results of Fig. \ref{fig03}, within the experimental accuracy, a bigger increase of the  dynamical emission probability  is found when switching to a neutron rich target (from Zn to Ni) than when increasing the projectile $N/Z$ (from Xe to Sn): it can be inferred that the dynamical IMF emission probability, in this case of a heavy projectile on a lighter target, shows a greater sensitivity on the target $N/Z$ content rather than on the projectile one.
From the results of the 3 isobaric systems, it follows that dynamical IMFs emission probability grows with the $N/Z$ content of, both, target and projectile. Clearly, this agrees with the results for the two Sn+Ni systems, already seen in \cite{Rus15}, where Isospin content of both projectile and target changes; however this couple of systems, having the same coulomb effects, allowed us to suggest, in the previous mentioned work \cite{Rus15}, that the observed differences in dynamical IMFs emission probability had to be triggered by size or Isospin difference; now, the present analysis on isobaric systems excludes, within the experimental accuracy, any size effect.

\begin{figure}[h]
\includegraphics[width=1.0\textwidth]{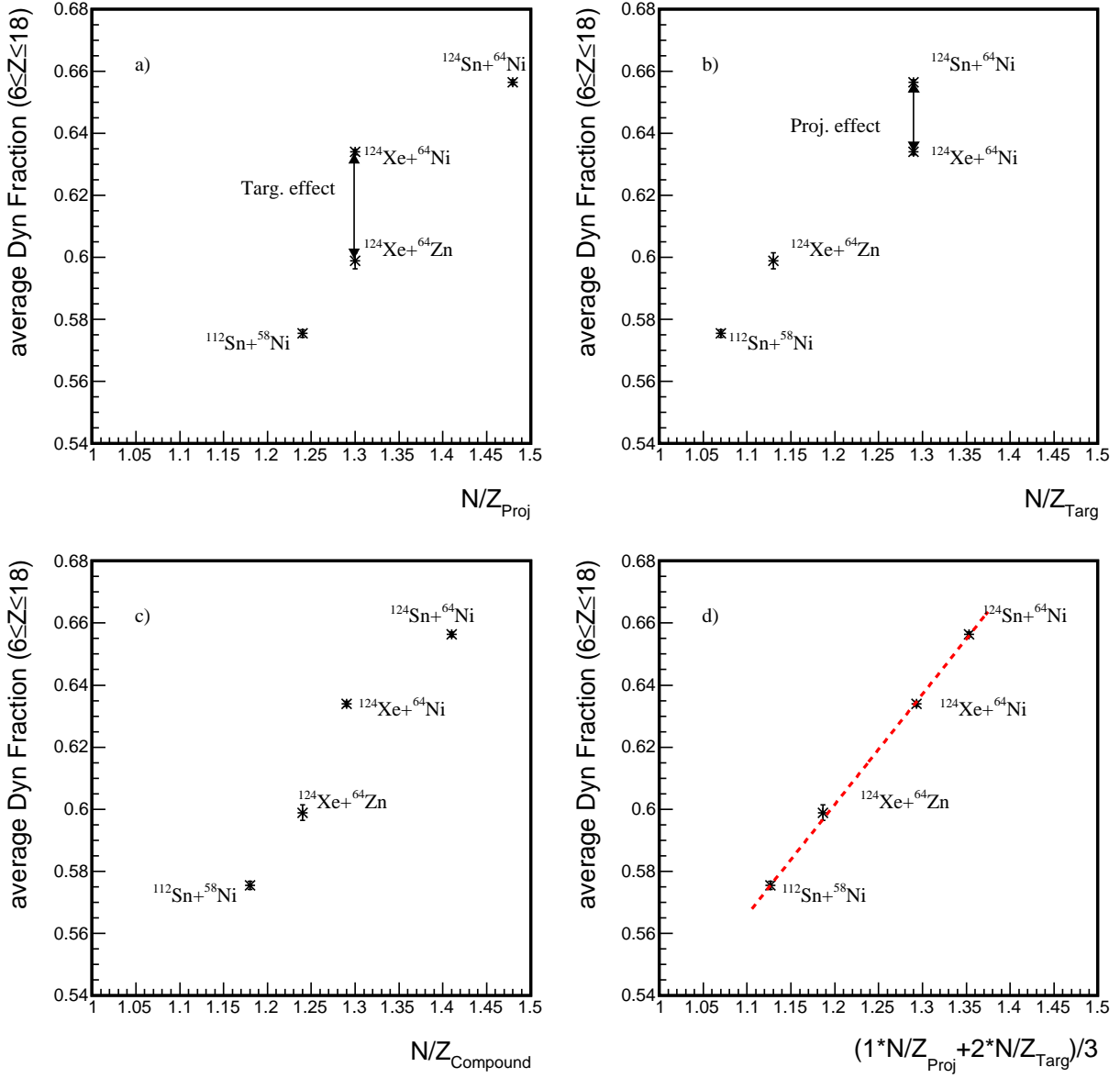}
\caption{ Weighted mean of the Dynamical emission probability for $6\leq Z \leq 18$, as extracted  from Fig. \ref{fig04}, vs the $N/Z$ content of projectile, target and compound system in panels a), b) and c), respectively, and the value $(1\times(N/Z)_{Proj}+2\times(N/Z)_{Targ})/3$ in panel d). Dotted red line in panel d) is obtained by least-squares method.
\label{fig04bis}}
\end{figure}

It is important to notice and highlight that in \cite{Rus15} we were able to extract the cross section associated to dynamical and statistical emission for the two Sn+Ni systems. By comparing that (see Fig. 4 of Ref. \cite{Rus15}), it resulted that while the ratio of the statistical cross sections stays approximately constant, as a function of Z, at a value of about 1.1, the ratio of the dynamical cross sections steady increases with increasing Z of the IMF, from about 1.3 for light IMFs to about 2.0 for the heaviest IMFs. This quantitative evaluation allowed us to say that the enhanced IMFs emission, observed increasing the $N/Z$ content, was essentially due to an increase of the dynamical emission contribution.\\
We can also notice that, for the lighter IMFs (Z$\lesssim$6) no appreciable difference is found among the studied systems. 
In contrast with the heavier fragments, dynamically produced in a binary-like PLF* splitting, light IMFs are produced by the fragmentation of a transient short-living ($\sim 10-100 fm/c$) low density neck-region, reminiscent of the partial overlap between the projectile and target \cite{Def05a,Def14}. Density dependence of the clustering process in the very early phase of the collision was found consistent with transport simulations \cite{Def12,Def14}.
The observed difference between Xe and Sn projectile for lithium fragment is due essentially to the fact that, in the new data, a slightly higher triggering threshold in the energy released in the silicon detector ($\Delta E$ energy loss) was set, resulting in loss of the fast Z=3 fragments.\\ 
At this point, we consider useful to quantitatively  investigate the correlation between the observed Dynamical emission probability and the $N/Z$ content. For that, we have plotted the weighted mean of the Dynamical emission probability for $6\leq Z \leq 18$, as extracted  from Fig. \ref{fig04}, vs the $N/Z$ content of projectile, target and compound system in panels a), b) and c) of Fig. \ref{fig04bis}, respectively. In all the three panels we can observe the increase of the Dynamical emission probability with the $N/Z$ content. In panels a) and b) also the effect of changing only the  $N/Z$ content of the target, or projectile, alone is evidenced; from panel a), when going from $^{124}Xe+^{64}Ni$  to $^{124}Sn+^{64}Ni$, the increase of the $N/Z$ of projecile by $\sim 14\%$  leads to an increase of Dynamical emission probability of $\sim 3\%$;  from panel b), when going from $^{124}Xe+^{64}Zn$ to $^{124}Xe+^{64}Ni$, the increase of the $N/Z$ of target by $\sim 14\%$  leads to an increase of Dynamical emission probability of $\sim 6\%$; this confirms what discussed above, the greater sensitivity to the  $N/Z$ content of the target. We can also notice that there is no strict linear correlation in panel c). However, in panel d) we present the correlation between a weighted $N/Z$ content, $(1\times(N/Z)_{Proj}+2\times(N/Z)_{Targ})/3$ and the Dynamical emission probability. With such a choice,  a linear relation is found, as indicated by the least-squares dotted red line. This can be interpreted as a quantitative indication that dynamical emission probability is twice more sensitive to the  target $(N/Z)$ content with respect to the projectile one.\\
\section{Comparison with a statistical-evaporative code}
\begin{figure}[h]
\includegraphics[width=1.0\textwidth]{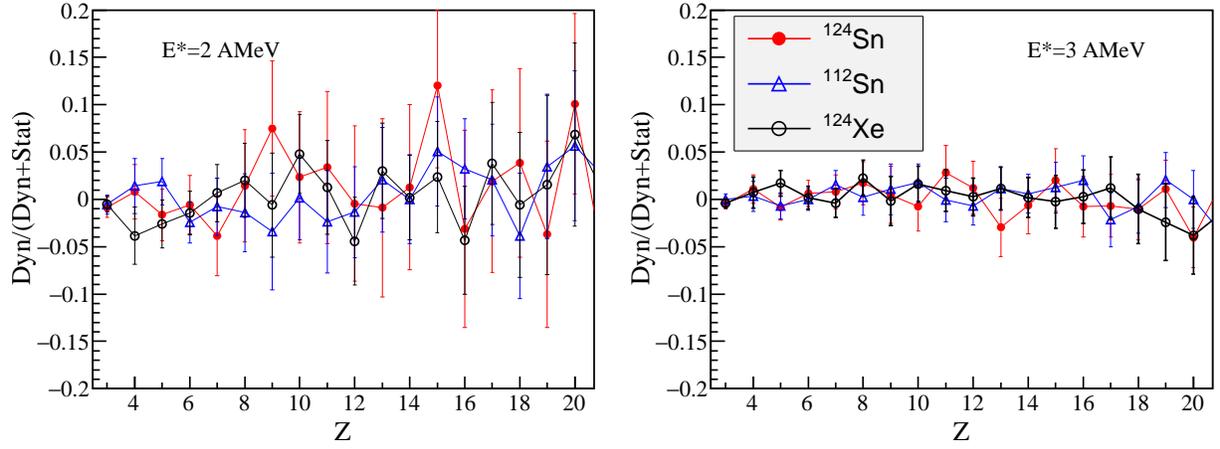}
\caption{(Color online) Ratio of the dynamical component to the total (dynamical+statistical) value, plotted as
a function of the IMF atomic number $Z$,  obtained simulating the projectile de-excitation process, for the 4 systems of fig. \ref{fig04}, using the Gemini++ statistical model and assuming an excitation energy of 2 AMeV (left panel) and 3 AMeV (right panel). 
\label{fig06}}
\end{figure}

In order to verify that the procedure to calculate the probability associated to Dynamical emission is robust and not biased by any effects of equilibrated secondary decays simply related to the different N/Z content of the projectile-systems here investigated, we have performed calculations using the statistical-evaporative code, Gemini++ \cite{Cha10,Man10}. We have simulated $10^{6}$ de-excitation events from $^{124}Sn$, $^{112}Sn$ and $^{124}Xe$ projectile-nuclei with excitation energy of 2 or 3 AMeV and fixed angular momentum of 30 $\hbar$ \cite{Def05b}. We have then applied the same selection and procedure used during the data analysis to the fragments produced by the model. The obtained ratio of the dynamical component, as definde by the $\cos(\theta_{prox})$ method, to the total one is shown in the two panels of Fig. \ref{fig06} for 2 MeV and 3 MeV of excitation energy, left and right panel, respectively. Apart for the statistical error, as expected the probability of dynamical component is null, whatever the charge of the IMFs, probing the fairness of our analysis scheme and assumptions. 


\section{CONCLUSIONS}

The IMFs production probability from projectile-like fragmentation in semi-peripheral reactions has been studied  in collisions of $^{124}$Xe projectiles with two different targets of $^{64}$Ni and $^{64}$Zn at the laboratory energy of 35 \amev. The two systems, having the same size (isobaric), differs in the target Isospin ($N/Z$).
The data analysis confirms the coexistence  of dynamical and statistical behaviors in the PLF* emission mechanism, as already observed for Sn projectiles \cite{Rus15}. Probability of the prompt-dynamical emission of IMFs is enhanced in the case of neutron rich $^{64}$Ni target. 
Through a comparison with previous data for the $^{124,112}Sn+^{64,58}Ni$ systems, it is experimentally investigated the sensitivity of the prompt-dynamical component of the PLF* decay on both the projectile an target Isospin $N/Z$ content. It is found that this dynamical emission is favored by an increase of both, and separately, projectile and target Isospin, and, in this case of a heavy projectile on a lighter target, shows an higher sensitivity to the target Isospin content. In addition, the influence of the system size on inducing dynamical  IMFs emission can be excluded.

 This observation allows to highlight the role of Isospin content in the growth 
of those dynamical instabilities, taking place during the early interaction and re-separation phases, triggering the dynamical IMFs emission mechanism here studied.
To shed lights on the interconnection between these results and properties of the many body microscopic interaction, calculations with dynamical transport models able to careful describe the projectile-target interaction and re-separation phases and to follow the full time path (dynamical and equilibrated) of the excited PLF* are needed.

It follows that the present analysis and our recent investigations \cite{Pag04,Def05a,Def05b,Rus10,Def12,Def14,Rus15}, trying to study fragment production in a large range of time scales, from fast-dynamical fragmentation of the dilute neck-region to the equilibrated de-excitation at normal nuclear density, and for the whole IMFs mass spectrum, represent useful constraints for reaction simulations \cite{Col98,Pap07,Bar05,BaLi08}.

\begin{acknowledgments}
We thank the INFN-LNS staff for providing both beams and targets of excellent quality. We are grateful also to the electronics, advanced technology and mechanical design staffs of INFN-division of Catania. 
\end{acknowledgments}
\bibliography{inter015_4.bib}

\begin{thebibliography}{31}
\bibitem [{Not()}]{Not}%
  \BibitemOpen
\bibfield  {journal} {  }\href@noop {} {}\bibinfo {note} {For the sake of
  simplicity, we assume that in a semi-peripheral collision of a projectile P
  and a target T, the binary (at first) system PLF* + TLF* is formed,
  $P+T\rightarrow PLF^{*}+TLF^{*}$. A two-step-decay scenario is assumed: The
  IMF is emitted from the projectile primary fragment PLF* (or TLF*),
  $PLF^{*}\rightarrow PLF+IMF$. In the analysys, the PLF* is reconstructed from
  the PLF-IMF pair, neglecting neutrons and Light Charged Particles ($Z\leq2$)
  emission.}\BibitemShut {Stop}%
\end{thebibliography}

\end{document}